\documentclass[journal, twoside]{IEEEtran}
\usepackage{cite}
\usepackage{balance}
\usepackage{multirow}

\ifCLASSINFOpdf
\else
\fi
\usepackage{amssymb}
\usepackage{array}
\usepackage{amsfonts}   

\usepackage[dvips]{graphicx}

\usepackage[cmex10]{amsmath}
\usepackage[tight,footnotesize]{subfigure}
\usepackage{threeparttable}
\usepackage{ulem}
\newcommand{\superscript}[1]{\ensuremath{^{\textrm{#1}}}} 

\begin{document}
%
\title{A Novel Data-Aided Channel Estimation with Reduced Complexity for TDS-OFDM Systems}
%
%
%

\author{Ming~Liu, 
        Matthieu ~Crussi\`ere,~\IEEEmembership{Member,~IEEE,}
        and~Jean-Fran\c{c}ois~H\'elard,~\IEEEmembership{Senior~Member,~IEEE}
\thanks{Authors are with Universit\'e Europ\'eenne de Bretagne (UEB), INSA, IETR, UMR 6164, 20 avenue des Buttes de Co\"esmes, F-35708 Rennes, France.
 e-mail: \{ming.liu, matthieu.crussiere, jean-francois.helard\}@insa-rennes.fr.}
}

\maketitle

\begin{abstract}
In contrast to the classical cyclic prefix (CP)-OFDM, the time domain synchronous (TDS)-OFDM employs a known pseudo noise (PN) sequence as guard interval (GI). Conventional channel estimation methods for TDS-OFDM are based on the exploitation of the PN sequence and consequently suffer from intersymbol interference (ISI). This paper proposes a novel
data-aided channel estimation method which combines the channel estimates obtained from the PN sequence and, most importantly, additional channel estimates extracted from OFDM data symbols.
Data-aided channel estimation is carried out using the rebuilt OFDM data symbols as virtual training sequences. In contrast to the classical turbo channel estimation, interleaving and decoding functions are not included in the feedback loop when rebuilding OFDM data symbols thereby reducing the complexity.
Several improved techniques are proposed to refine the data-aided channel estimates, namely one-dimensional (1-D)/two-dimensional (2-D) moving average and Wiener filtering.
Finally, the MMSE criteria is used to obtain the best combination results and an iterative process is proposed to progressively refine the estimation. Both MSE and BER simulations using specifications of the DTMB system are carried out to prove the effectiveness of the proposed algorithm even in very harsh channel conditions such as in the single frequency network (SFN) case.
\end{abstract}


%
\IEEEpeerreviewmaketitle

\section{Introduction}

\IEEEPARstart{C}{onventionally}, orthogonal frequency division multiplexing (OFDM) adopts cyclic prefix (CP) as guard interval (GI) to mitigate intersymbol interference (ISI) and enable simple equalization at receiver side. However, CP is a redundant copy of the data, which reduces the useful bit-rate and degrades the spectrum efficiency. Recently, it has been proposed to replace CP by a known pseudo-noise (PN) sequence. The main advantage of this approach is that the known sequence, which initially serves as GI, can be reused as training sequence to carry out channel estimation and synchronization.
Therefore no pilot is inserted among the useful data symbols in the frequency domain as needed in traditional CP-OFDM.
This kind of OFDM waveform is referred to as time domain synchronous-OFDM (TDS-OFDM)\footnote{It is also known as pseudo random postfix-OFDM (PRP-OFDM)~\cite{PRPOFDM} and known symbols padding-OFDM (KSP-OFDM)~\cite{KSPOFDM} in other literatures.}~\cite{WANG05} and has been adopted by the Chinese digital television/terrestrial multimedia broadcasting (DTMB) system~\cite{DTMB,DTMB_single}. After perfectly removing the PN sequence, the received TDS-OFDM signal is converted to the so-called zero padding (ZP)-OFDM~\cite{ZPOFDM}. Then, all the well established equalization techniques~\cite{ZPOFDM,WANG06} for ZP-OFDM can be applied for TDS-OFDM signals.

The usage of the PN as GI is somehow ``double-edged sword''. On one hand, it improves the spectrum efficiency by economizing the cost of transmitting redundant data in the CP and the pilots in the data symbols; on the other hand, the channel memory effect introduces mutual interference between the PN sequence and the OFDM data symbols. That is to say, the interference from data symbols compromises  the performance of the PN-sequence-based channel estimation  and  the inaccurate channel estimate consequently causes imperfect PN sequence removal and further increases the difficulty to recover the data symbols. The channel estimation is thus the key point in the TDS-OFDM based transmissions.

\begin{figure*}[!t] 
\centering
\includegraphics[width=18cm]{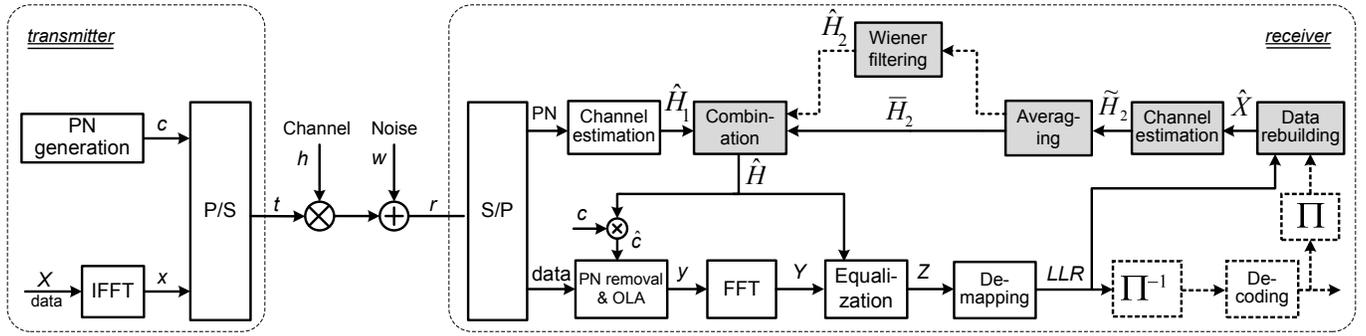}
\caption{Block diagram of TDS-OFDM system. The shaded blocks are the processing used for the proposed method. The dashed blocks are the additional processing needed for the turbo channel estimation.}
\label{fig_TDS_OFDM_baseband}
\end{figure*}

Many papers have addressed this challenging issue in the past few years and can be classified into three categories:

1. \textit{Making channel estimation  based only on the PN sequence.} In~\cite{SONG05}, the time domain cross-correlation between  the received PN sequence and the locally generated one is used to make channel estimation. It exploits the correlation property of the PN sequence in order to obtain  processing gain to combat the noise and ISI. The complexity significantly grows with the increase of the length of the channel delay spread. In~\cite{PRPOFDM}, a simple channel estimator is proposed based on the expectation (mean value) of the received PN sequence which cancels the random ISI component. However, this algorithm only adapts to the static channel situations and requires large storage.

2. \textit{Removing ISI from the PN sequence when making channel estimation.} A first approach~\cite{WANG05} proposed to iteratively remove the ISI using the decision feedback method. However, much complexity is spent by a series of fast Fourier transforms (FFT).
In~\cite{TANG07}, it is proposed to carry out the channel estimation using the PN sequence and its tail spread in the beginning part of each OFDM data symbol. By iteratively removing the data symbol, the channel estimation is progressively improved. The estimation results however suffer an error floor even at low signal to noise ratio (SNR) due to the existence of  ISI in the PN sequence.
\cite{TANG08} proposed to alternately remove the mutual interference between the PN sequence and the OFDM data symbols. A so-called partial decision is also used to reduce the error detection of data symbols.
However, the performance significantly degrades when the channel delay spread is long.
\cite{YANG08TCE} also adopted the partial decision technique to remove the interference on the PN sequence from previous data symbols. In order to avoid the interference from the following data symbols, the ``tail'' of the received PN sequence is estimated and then used to reconstruct the interference-free received PN sequence for channel estimation.

3. \textit{Using the pilots in a complementary way to the PN sequence to enhance the channel estimation.} It is proposed to replace some data subcarriers by pilots to improve the channel estimation performance in~\cite{KSPOFDM}. However, the spectrum efficiency is compromised.  Another possible solution is the data-aided channel estimation method which uses the rebuilt data symbols as ``virtual pilots'' to make channel estimation and thus does not need any extra pilots. In~\cite{YANG08}, it is proposed to make hard decisions to the equalized symbols that fall in the reliable decision region. The hard decided data symbols are used as virtual pilots for channel estimation.
However, this method requires an additional PN frame header for several OFDM symbol to obtain a good initial channel estimate. Therefore, it is not straightforwardly applicable to the existing DTMB system.
In~\cite{ZHAO08}, a turbo channel estimation algorithm is proposed  in the context of CP-OFDM. An obvious disadvantage of this method is the extremely high complexity and long time delay, especially in the systems with sophisticated  channel decoder and deep interleaver. For example, the DTMB system adopts LDPC codes and convolutional interleaver with a time delay of 170 (or even 510) OFDM frames which prohibits the using of such a method.

The aim of this paper is to propose a low-complex but effective algorithm to increase the robustness of the channel estimation in TDS-OFDM compared to the already existing techniques described above.
The highlights of algorithm proposed in this paper are (a) excluding the channel decoder, interleaver and de-interleaver from the iterative channel estimation process to reduce complexity,
(b) rebuilding \textit{soft symbols} for channel estimation to prevent error propagation,
(c) exploiting the correlation property of the channel to obtain improved estimation results, and
(d) minimum mean square error (MMSE) combination of channel estimates obtained from PN and data.

More concretely, we propose to rebuild the data symbols using the likelihood information from the \textit{demapper}. Then the instantaneous data-aided channel estimates are acquired using these rebuilt data symbols. As the interleaving and channel decoding processes are not included in the feedback loop, the computational complexity is significantly reduced compared to the turbo channel estimation. The \textit{soft symbols} are fed back in the iterations serving as virtual pilots in the channel estimation. Since no decision is made on the rebuilt symbols, the proposed method does not cause error propagation due to imprudent hard decisions. The uncertainty (noise) is kept in the soft data symbols and in the resulting data-aided channel estimates as well. Moreover, we propose several improved techniques, namely one-dimensional (1-D)/two-dimensional (2-D) moving average and Wiener filtering, which exploit the time-frequency correlation property of the channel to suppress the noise existing in the data-aided channel estimates. The cooperation of these refining techniques and soft symbol rebuilding achieves satisfactory channel estimation accuracy with low computational complexity. The final channel estimates combine the results obtained from PN and from data according to MMSE criteria. The proposed method can adapt to difficult situations, including higher order constellations, extremely long channel spread and channel time variations.

The rest of the paper is organized as follows. In section II, the mobile channel and TDS-OFDM signal models are described. In section III, the PN based channel estimation is presented. The new proposed data-aided channel estimation method is presented in section IV. Computational complexity is evaluated in section V. Simulation results are shown in section VI. Conclusions are drawn in section VII.

In this paper, $(\cdot)^*$ denotes the conjugate of complex number, $\mathbb{E}\{\cdot\}$ is the expected value, $(\cdot)^T$ and $(\cdot)^H$ are the matrix transpose and Hermitian transpose, respectively.

\section{System Model}
\subsection{Discrete Channel Model for OFDM}

The wireless channel is modeled as an $L$\superscript{th} order time-varying  finite impulse response (FIR) filter. The parameter $L$ is determined by the maximum excess delay of the channel. The  channel is assumed to be quasi-static, namely channel coefficients remain constant within one OFDM symbol duration but change from one OFDM symbol to another. The channel impulse response (CIR) for the $i$\superscript{th} OFDM symbol is:
\begin{equation}
  h[i,m]=\sum_{l=0}^{L-1}h^{(i)}_l\delta[m-l],
\end{equation}
where $\delta[\cdot]$ is the Kronecker delta function, $h_l^{(i)}$ is the $l$\superscript{th} filter tap and is modeled as zero mean complex Gaussian random variables. Furthermore, the channel filter taps are assumed to follow the wide sense stationary uncorrelated scattering (WSSUS) assumption, namely different taps are statistically independent, while a specific tap is correlated in time. The channel frequency response (CFR) for the $k$\superscript{th} subcarrier of the $i$\superscript{th} OFDM symbol can be obtained via an $N$-point FFT over the CIR:
\begin{equation}
  H[i,k]=\sum_{l=0}^{N-1}h_l^{(i)}e^{-j\frac{2\pi}{N}lk}.
\end{equation}

The time-frequency correlation function of the CFR is \cite{YeLI00}:
\begin{equation}
  \phi_H[p,q]\triangleq \mathbb{E}\left\{H[i+p, k+q]H^{\ast}[i,k]\right\}= r_t[p]r_f[q],
\end{equation}
where $r_t[p]$ and $r_f[q]$ are the time and frequency domain correlation functions of the CFR, respectively. Concretely, taking the Jakes' mobile channel model, $r_t[p]$ is~\cite{YeLI00}:
\begin{equation}
\label{eqn_corr_time}
  r_t[p] = J_0(2\pi p f_dT_b),
\end{equation}
where $J_0(\cdot)$ is the zero-order Bessel function of the first kind, and $f_{d}$ is the maximum Doppler frequency related to velocity $v$ and carrier frequency $f_c$ by $f_d = vf_c/c$, where $c$ is the speed of light, $T_b$ is the time duration of one OFDM block and $T_b=T_g+T$ given $T_g$ and $T$ the durations of GI and OFDM data parts, respectively. The frequency domain correlation $r_f[q]$ is:
\begin{equation}
\label{eqn_corr_freq}
  r_f[q]= \sum_{l=0}^{L-1}\sigma^2_{l}e^{-j\frac{2\pi}{N} ql}.
\end{equation}
where $\sigma_l^2$ is the power of the $l$\superscript{th} path.  Without loss of generality, the power of the channel is normalized so that $\sum_{l=0}^{L-1}\sigma_l^2=1$.

\begin{figure}[!t] 
  \begin{center}
    \includegraphics[width=3.5in]{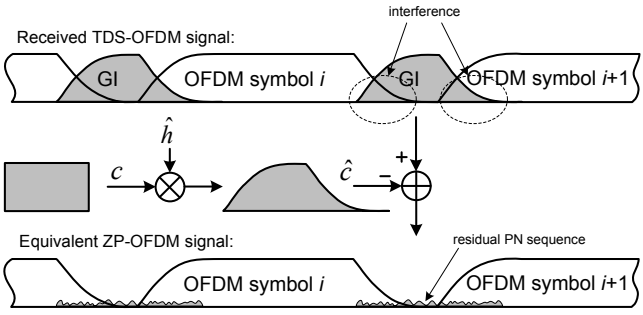}
    \caption{PN removal of the TDS-OFDM signal.}
    \label{fig_PN_removal}
  \end{center}
\end{figure}

\subsection{TDS-OFDM Signal Model}
Fig. \ref{fig_TDS_OFDM_baseband} presents the baseband model of the TDS-OFDM based system. The $i$\superscript{th} OFDM data symbol is formed by $N$-point inverse fast Fourier transform (IFFT):
\begin{equation}
  \label{eqn_X}
    x^{(i)}[n] = \frac{1}{\sqrt{N}}\sum_{k=0}^{N-1} X[i,k]e^{j\frac{2\pi}{N}nk}, 0\leq n\leq N-1.
\end{equation}

A $\nu$-length PN sequence $\{c^{(i)}[n]\}_{n=-\nu}^{-1}$ is then inserted before $\{x^{(i)}[n]\}_{n=0}^{N-1}$ as GI. The $i$\superscript{th} transmitted time domain signal is thus:
\begin{equation}
\label{eqn_time_domain_sig}
    t^{(i)}[n]= \left\{{\begin{array}{*{20}c} {c^{(i)}[n]} & {-\nu \leq n < 0} \\
   {x^{(i)}[n]} & {0 \leq n \leq N-1}  \\\end{array}} \right. .
\end{equation}

After passing the multipath fading channel, the received signal is a linear convolution of the transmitted
signal and the channel as given in (\ref{eqn_linear_conv_detail}) which is shown in the top of next page,

\newcounter{MYtempeqncnt}
\begin{figure*}[!t]
\normalsize
\setcounter{MYtempeqncnt}{\value{equation}}
\setcounter{equation}{7}
\begin{eqnarray}
\label{eqn_linear_conv_detail}
   &&r^{(i)}[n]=\sum_{l=0}^{L-1}h^{(i)}_lt^{(i)}[n-l]+w[n]\nonumber\\
  &=&\left\{{\begin{array}{*{50}l}
\sum_{l=0}^{n+\nu}h_l^{(i)}c^{(i)}[n-{l}] + \sum_{l=n+\nu+1}^{L-1}h_l^{(i)}x^{(i-1)}[n-l]_N  + w[n]& {-\nu \leq n < -\nu +L-1}\\\noalign{\vskip4pt}
   \sum_{l=0}^{L-1}h_l^{(i)}c^{(i)}[n-l] + w[n] & {-\nu+L-1 \leq n < 0}  \\\noalign{\vskip4pt}
\sum_{l=0}^{ n}h_l^{(i)}x^{(i)}[n-l]
+ \sum_{l=n+\nu+1}^{L-1}h_l^{(i)}c^{(i)}[n-l] + w[n]& 0\leq n < L-1\\\noalign{\vskip4pt}
   \sum_{l=0}^{L-1}h_l^{(i)}x^{(i)}[n-l] + w[n] &   L-1\leq n < N\\\end{array}} \right.
\end{eqnarray}
\setcounter{equation}{8}
\hrulefill
\end{figure*}
where $[n]_N$ is the residue of $n$ modulo $N$ and $w$ is the additive white Gaussian noise (AWGN) with a variance of $\sigma^2_{w}$.
From (\ref{eqn_linear_conv_detail}) and Fig. \ref{fig_PN_removal}, it can be found that there are some mutual interference between the OFDM data symbols and the PN sequences due to the channel memory. In order to recover the data symbols, it is necessary to remove the PN sequence from the received signal first. Since the PN sequence $c^{(i)}[n]$ is perfectly known by the receiver, the channel-distorted PN sequence can be estimated by making a linear convolution of PN sequence $c^{(i)}[n]$ and the estimated channel response $\hat h^{(i)}$:
\begin{equation}
\label{eqn_est_pn}
  \hat c^{(i)}[n] = \sum_{l=0}^{L-1}\hat h_l^{(i)}c^{(i)}[n-l], \quad -\nu \leq n <L-1.
\end{equation}

Using $\hat c^{(i)}[n]$, the PN sequence is removed from the received signal as depicted in Fig. \ref{fig_PN_removal}. If the channel is perfectly estimated, the PN sequence and its tail can be completely removed from the received signal. Otherwise, there will be some residual contribution of the PN sequence in the received signal. The PN sequence removal is expressed in (\ref{eqn_received_sig_after_pn_rml}) which is given in the next page, where $\Delta h_l^{(i)}=h_l^{(i)} - \hat h_l^{(i)}$ is the estimation error of the $l$\superscript{th} channel tap.

\begin{figure}[!t]
  \begin{center}
    \includegraphics[width=3.5in]{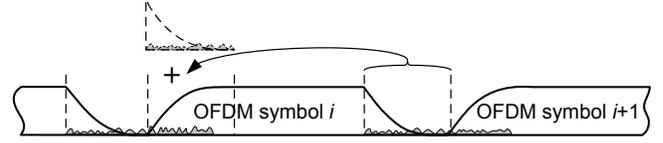}
    \caption{Illustration of the OLA process.}
    \label{fig_OLA_noisy}
  \end{center}
\end{figure}

\newcounter{MYtempeqncnt1}
\begin{figure*}[!t]
\normalsize
\setcounter{MYtempeqncnt1}{\value{equation}}
\setcounter{equation}{9}
\begin{eqnarray}
\label{eqn_received_sig_after_pn_rml}
   &&\bar{r}^{(i)}[n]= r^{(i)}[n] - \hat c^{(i)}[n] \nonumber\\
   &=&\left\{{\begin{array}{*{50}l}
\sum_{l=0}^{n+\nu}\Delta h_l^{(i)}c^{(i)}[n-l] + \sum_{l=n+1}^{L-1}h_l^{(i)}x^{(i-1)}[n-l]_N  + w[n]& {-\nu \leq n < -\nu +L-1} \\\noalign{\vskip4pt}
   \sum_{l=0}^{L-1}\Delta h_l^{(i)}c^{(i)}[n-l] + w[n] & {-\nu+L-1 \leq n < 0}  \\\noalign{\vskip4pt}
\sum_{l=0}^{ n}h_l^{(i)}x^{(i)}[n-l]
+ \sum_{l=n+1}^{L-1}\Delta h_l^{(i)}c^{(i)}[n-l] + w[n]& 0\leq n < L-1\\\noalign{\vskip4pt}
   \sum_{l=0}^{L-1}h_l^{(i)}x^{(i)}[n-l] + w[n]&   L-1\leq n < N\\\end{array}} \right.
\end{eqnarray}
\setcounter{equation}{10}
\hrulefill
\end{figure*}

After removing the PN sequence, the TDS-OFDM signal is turned to an equivalent ZP-OFDM signal. Thus, the so-called overlap and add (OLA) \cite{KSPOFDM} process is performed by adding the following GI to the beginning part of the OFDM symbol as shown in Fig. \ref{fig_OLA_noisy}. More precisely, the OLA process can be written as:
\begin{equation}
\label{eqn_ola}
  y^{(i)}(n)= \bar{r}^{(i)}(n) + \bar{r}^{(i+1)}(n-\nu),\; 0\leq n < L-1.
\end{equation}
Suppose that the length of GI exceeds that of the channel, there is no ISI between two adjacent OFDM symbols. Taking (\ref{eqn_received_sig_after_pn_rml}) into account, the received signal after removing the GI is written as:
\begin{align}
    \label{eqn_linear_circ_conv}
    r^{(i)}[n]&\!=\!\!\!\sum_{l=0}^{L-1}h_l^{(i)}x^{(i)}[n-l]_N \!+\! w'[n] \!+\!\xi [n],0\leq n<N,
\end{align}
where $w'(n)$ is the noise after OLA which is slightly colored and boosted in the OLA process. The equivalent noise power is $\sigma_{w'}^2=\frac{N+\nu}{N}\sigma^2_{w}$ \cite{KSPOFDM}. $\xi[n]$ is the contribution of the residual PN sequence in the received signal. From (\ref{eqn_linear_circ_conv}), it can be seen that the linear convolution of the channel and data becomes a circular one after OLA process. Therefore, after fast Fourier transform (FFT), the received frequency domain TDS-OFDM signal has finally a similar representation as CP-OFDM:
\begin{equation}
  Y[i,k]\!\!=\!\!\frac{1}{\sqrt{N}}\!\!\!\sum_{n=0}^{N-1}\!\!y^{(i)}
  \![n]e^{-j\frac{2\pi}{N}nk}\!\!\!=\!H[i,k]X[i,k]\!+\!\!W'[i,k]
\end{equation}
where $W'$ is AWGN with the same variance as $w'$. The TDS-OFDM signal can thus be easily equalized by a one-tap equalizer:
\begin{equation}
  Z[i,k] = \frac{Y[i,k]}{H[i,k]}.
\end{equation}

\section{PN-based Channel Estimation}
As specified in~\cite{DTMB}, the $\nu$-length PN sequence in the GI is composed of an $N_{\rm PN}$-length PN sequence, more specifically a maximum-length sequence (m-sequence), as well as its pre- and post-circular extensions.
Since any shift of an m-sequence is itself an m-sequence, the GI can also be treated as another $N_{\rm PN}$-length m-sequence with its CP.
If this CP is longer than the length of CIR, the $N_{\rm PN}$-length PN sequence is ISI-free. Using the ISI-free PN sequence, a least square (LS) channel estimation  is made for the  $i$\superscript{th} OFDM symbol:
\begin{equation}
 \label{eqn_initial_chest}
  \bar{H}_1[i,k] = \frac{S[i,k]}{P[i,k]}=H[i,k]+\frac{W[i,k]}{P[i,k]},0\leq k < N_{\rm PN},
\end{equation}
where $P$ and $S$ are obtained through $N_{\rm PN}$-point FFT applied on the transmitted and received ISI-free PN sequences, respectively. The CIR estimate is
\begin{equation}
    \hat h[i,l] = \frac{1}{N_{\rm PN}}\sum_{k=0}^{N_{\rm PN}-1}\bar{H}_1[i,k]e^{j\frac{2\pi}{N_{\rm PN}}kl}.
\end{equation}
Consequently, the $N$-length CFR estimation $\hat H_1$ is obtained by applying $N$-point FFT on $\hat h[i,l]$. The mean square error of $\hat H_1$ is:
\begin{equation}
\label{eqn_mse_pn_est}
    \varepsilon_{\hat H_1}\!\!\!=\!\!\frac{1}{N}\!\!\!\sum_{k=0}^{N-1}\!\mathbb{E}\!\left\{\!|H[i,k] \!-\! \hat H_1[i,k]|^2\!\right\}\!\!=\!\!\frac{L \sigma_w^{2}}{N_{\rm PN}}\!\!\sum_{k=0}^{N_{\rm PN}-1}\!\!\!\!\frac{1}{|P[i,k]|^2}.
\end{equation}
As mentioned before, the channel estimation error results in interference to the OFDM data symbols. From the analysis given in the appendix, the power of the interference is computed as:
\begin{eqnarray}
  \sigma_I^2[k]&=&\frac{1}{N}\sum_{l=0}^{L-1}\sigma_{\Delta h_l}^2 \left[\  \sum_{n=0}^{\nu-1}|c[n]|^2\right. \nonumber\\
&+&\!\!\left. \sum_{q=1}^{\nu-1}2\cos\left(\frac{2\pi}{N}kq\right)\!\!\sum_{n=0}^{\nu-1-q}c_l[n]^{\ast}c_l^{\ }[n+q]\right].
\end{eqnarray}
The part in the bracket of the above equation is only determined by the PN sequence and is constant for a given sequence. Thus, the power of the interference is determined by the length of channel $L$ and the MSE of CIR estimation. Hence, the worse the channel estimation, the more interference will be introduced to the OFDM data symbols, which motivates us to propose additional processing to improve the channel estimation results.

\section{Data-aided Channel Estimation}
\subsection{Instantaneous Data-aided Channel Estimate}
In contrast to the classical turbo channel estimation algorithm like~\cite{ZHAO08}, the new proposed method \textit{excludes} the deinterleaving, channel decoding  and  interleaving processings from the feedback loop as shown in Fig. \ref{fig_TDS_OFDM_baseband}. In other words, the soft data symbols used for data-aided channel estimation are rebuilt using the likelihood information from the \textit{demapper}.

The soft-output demapper  demodulates the complex data symbols into Log-likelihood ratio (LLR) of bits \cite{LLR}. The LLR $\lambda_l[i,k]$ corresponding to the $l$\superscript{th} bit of the $(i,k)$th equalized data symbol $Z[i,k]$ is defined as:
\begin{eqnarray}
\label{eqn_prb_bits}
    \lambda_l[i,k] & \triangleq & \log \frac{P\left(b[i,k,l]=1 | Z[i,k]\right)}{P\left(b[i,k,l]=0 | Z[i,k]\right)},
\end{eqnarray}
where the $P\left(b[i,k,l]=1 | Z[i,k]\right)$ is the conditional probability of the  $l$\superscript{th} bit equal to 1  given $Z[i,k]$. The sign of the LLR decides the corresponding bit equal to 1 or 0, and its absolute value gives the reliability of the decision. Based on the LLR, the probabilities of a bit equal to 1 and 0 are $P\left(b[i,k,l]=1\right) = \frac{e^{\lambda_l[i,k]}}{1+e^{\lambda_l[i,k]}}$ and $P\left(b[i,k,l]=0\right)=1-P\left(b[i,k,l]=1\right)$, respectively. Then they are  used as  \textit{a priori} probabilities to estimate the data symbols. First, the probability that the transmitted symbol $X[i,k]$ is equal to a specific constellation point $\alpha_j$ is computed as the product of the probabilities of all the bits belonging to this constellation:
\begin{equation}
\label{eqn_prb_constellation}
P\left( X[i,k]  = \alpha _j \right) = \prod_{l=1}^{\log_2 \mu} P\big( b[i,k,l]=\kappa_l(\alpha _j) \big),
\end{equation}
where $\Psi$ is the set of the constellation points of a given modulation scheme, $\mu$ is the modulation order and $\kappa_l(\alpha _j)\in \{0,1\}$ is the value of the $l$\superscript{th} bit of the constellation point $\alpha_j$. The soft data symbol is an expected value taking the \textit{a priori} probabilities (\ref{eqn_prb_constellation}) into account:
\begin{equation}
\label{eqn_data_rebuild}
\hat X[i,k] = \sum\limits_{\alpha _j  \in \Psi } {\alpha _j  \cdot P\left( { X[i,k]  = \alpha _j } \right)}.
\end{equation}

Note that, being different from the classical decision feedback channel estimation methods such as~\cite{TANG08} and~\cite{YeLI00}, no decision is made here in order to prevent  error propagation. Directly using the soft data symbols, an instantaneous channel estimate is obtained over all active subcarriers by:
\begin{eqnarray}
\label{eqn_scd_estimation}
&&\tilde{H}_2[i,k]=\frac{1}{\eta_{\hat X}[i,k]}\hat X[i,k]^{\ast} Y[i,k]\nonumber\\
&=&\!\!\!\!\!\!\frac{\hat X[i,k]^{\ast}X[i,k]}{\eta_{\hat X}[i,k]}H[i,k]\! + \! \frac{1}{\eta_{\hat X}[i,k]}\hat X[i,k]^{\ast}W''[i,k],
\end{eqnarray}
where $W''[i,k]$ is the noise and interference component with a variance $\sigma^{2}_{W^{''}}=\sigma^{2}_{W'}+\sigma^{2}_{I}$, $\eta_{\hat X}[i,k]=|\hat X[i,k]|^{2}$ is the power of the $(i,k)$th soft data symbol which is used as power normalization factor.
Note that, the power of the constellation is a constant and known value in the uniform power constellation cases such as BPSK or QPSK. Hence, the $\eta_{\hat X}[i,k]$ can be approximated by the power of the constellation $\eta_{\alpha}$ in order to reduce computational complexity.
If the data symbols are perfectly rebuilt, e.g. SNR is high, namely $\hat X[i,k]=X[i,k]$, (\ref{eqn_scd_estimation}) turns to an LS estimator:
\begin{equation}
\label{eqn_scd_est_simp}
\tilde{H}_2[i,k] = H[i,k] + \frac{W''[i,k]}{\sqrt{\eta_{X}[i,k]}}.
\end{equation}

Unlike the turbo channel estimation, the proposed method does not include any error correction before rebuilding the data. Therefore the rebuilt soft data symbols, i.e. $\hat X$'s, are affected by the noise as well as the channel fades. It is quite possible that the instantaneous channel estimate (\ref{eqn_scd_estimation}) is inaccurate and even erroneous for some subcarriers. Fortunately, as both the time delay spread and Doppler spectrum of the channel are limited, the CFR is highly correlated, i.e. almost identical within the coherence bandwidth and coherence time \cite{Rappaport}. Moreover, the coherence bandwidth spreads over several adjacent subcarriers in the OFDM system with large FFT size, and the coherence time is normally longer than several OFDM symbol durations in low and medium velocity cases. This enables us to refine the data-aided channel estimation using the correlation of the CFR. Several channel estimate refinement approaches are proposed in the following sections. According to the time-frequency range that data-aided approaches work, they are catalogued into 1-D (frequency domain) and 2-D (time-frequency domain) ones.

\subsection{1-D Refinements}
The proposed 1-D refinement approaches process all the instantaneous data-aided channel estimates within one OFDM symbol to get an improved estimation result at each time. Only the frequency domain correlation property of the channel is exploited. As the 1-D refinement approaches are carried out in a (OFDM) symbol-by-symbol fashion, they can track a fast variation of the channel and require minimum storage capacity. The OFDM symbol index $i$ is omitted in this section for the sake of notation simplicity.

\subsubsection{Moving Average}
Since the channel frequency response is almost identical within coherence bandwidth~\cite{Rappaport}, the most straightforward way to improve the data-aided estimation is to perform a moving average over the instantaneous channel estimates with a length less or equal to the coherence bandwidth. More specifically, the instantaneous channel estimates within a particular range, namely within a ``window'', are averaged to get a more precise estimate for the central position of the window. After the window sliding over all active subcarriers, a refined channel estimation is obtained. Concretely, the moving-averaged CFR for the $k$\superscript{th} subcarrier is expressed as:

\begin{eqnarray}
 \label{eqn_scd_esti_moving_average1D}
 &&\bar{H}_2[k]=\frac{1}{M}\sum_{m\in\Theta_k} \tilde{H_2}[m] \nonumber\\
 &\approx& \!\!\!\!\frac{H[k]}{M}\!\!\!\sum_{m\in\Theta_k}\!\!\!\frac{\hat X[m]^{\ast}X[m]}{\eta_{\hat X}[m]}\!+ \!\! \frac{1}{M}\!\!\!\sum_{m\in\Theta_k}\!\!\!\frac{\hat X[m]^{\ast}W''[m]}{\eta_{\hat X}[m]} \nonumber\\
 &\approx& \!\!\!\!H[k]\!+\!\frac{1}{M}\!\!\!\sum_{m\in\Theta_k}\!\!\!\frac{\hat X[m]^{\ast}W''[m]}{\eta_{\hat X}[m]}, 0\leq k \leq N-1,
\end{eqnarray}
where $\Theta_k = \{k-\lfloor\frac{M-1}{2}\rfloor, k-\lfloor\frac{M-1}{2}\rfloor+1,\ldots,k+\lfloor\frac{M-1}{2}\rfloor\}$ is the set of subcarrier indices within the moving average window with the $k$\superscript{th} subcarrier its central frequency, $M$ is the length of the moving average window and $W''$ is the noise and interference component. The moving average length $M$ is chosen less than or equal to the coherence bandwidth. It can be either empirically pre-selected according to the ``worst case'' or be adaptively chosen by computing the coherence bandwidth using the estimated CIR from initial PN based channel estimation. The variance of the data-aided CFR estimate for a specific subcarrier is:

\begin{eqnarray}
\label{eqn_mse_1D_averaged_sc}
 &&\sigma_{\bar{H}_2}^2[i,k]=\mathbb{E}\left\{\big|H[i,k]-\bar{H}_2[i,k]\big|^2\right\}\nonumber\\
 &=&\!\!\!\!\mathbb{E}\Big\{\Big|\frac{1}{M}\sum_{m\in\Theta_k}\frac{\hat X[i,m]W''[i,m]}{\eta_{\hat X}[i,m]}\Big|^2\Big\}\nonumber\\
 &=&\!\!\frac{\sigma_{W''}^2 }{M^2}\!\!\sum_{m\in\Theta_k}\frac{|\hat X[i,m]|^2}{\eta_{\hat X}[i,m]^2}= \!\!\frac{\sigma_{W''}^2 }{M^2}\!\!\sum_{m\in\Theta_k}\frac{1}{\eta_{\hat X}[i,m]}.
\end{eqnarray}
The MSE of the data-aided CFR estimate for the $i$\superscript{th} OFDM symbol can be estimated by averaging (\ref{eqn_mse_1D_averaged_sc}) over $N$ subcarriers:
\begin{eqnarray}
\label{eqn_mse_1D_averaged}
 &&\varepsilon_{\bar{H}_2,1D}=\frac{1}{N}\sum_{k=0}^{N-1}\mathbb{E}\left\{\big|H[i,k]-\bar{H}_2[i,k]\big|^2\right\}\nonumber\\
 &=&\!\frac{\sigma_{W^{''}}^2 }{NM^2} \sum_{k=0}^{N-1} \sum_{m\in\Theta_k} \frac{1}{\eta_{\hat X}[i,m]}.
\end{eqnarray}
Specifically, given the uniform power constellation with a power of $\eta_{\alpha}$,
\begin{equation}
\label{eqn_mse_sec}
\varepsilon_{\bar{H}_2,1D}= \! \frac{\sigma^2_{W^{''}}}{(M\eta_{\alpha})^2N}\!\sum_{k=0}^{N-1}\!\sum_{m\in\Theta_k}\!\!\!|\hat X[k] |^2\! =\! \frac{\sigma^2_{W^{''}}}{M{\eta_{\alpha}}^2}\bar{\eta}_{\hat X},
\end{equation}
where $\bar{\eta}_{\hat X}= \frac{1}{N}\sum_{k=0}^{N-1}|\hat X[k]|^2 $ is the mean power of the rebuilt data symbols.

\subsubsection{Wiener Filtering}
\begin{figure}[!t]
  \begin{center}
    \includegraphics[width=3.5in]{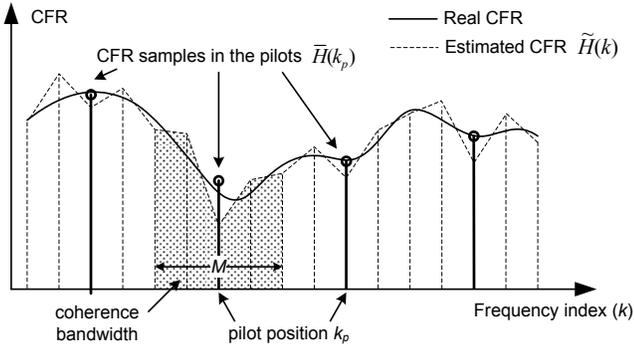}
    \caption{1-D averaging and interpolation.}
    \label{Fig_interpolate_1D}
  \end{center}
\end{figure}
The proposed Wiener filtering based refinement is depicted in Fig.\ref{Fig_interpolate_1D}. It consists of two steps: averaging and interpolation.

A bunch of subcarriers are first selected as ``virtual pilots''. The virtual pilots are equally spaced and the spacing $L_f$ (in terms of subcarriers)  is selected substantially fulfilling the sampling theorem~\cite{Kaiser}.  More concretely, with a $2\times$ oversampling ratio, $L_f$ is determined so that ${L_fL}/{N}\leq1/4$. Denote $\Xi$ the set of virtual pilots indices, i.e. $\Xi=\{k_p \big| k_p = p L_f, 0\leq p\leq \lfloor\frac{N}{L_f}\rfloor-1\}$. The cardinality of $\Xi$ (the quantity of virtual pilots) is $K_f$.

The instantaneous channel estimates within the coherence bandwidth around a virtual pilot $k_p\in\Xi$ are averaged to get a more accurate channel estimate $\bar{H}_2[k_p]$. Its expression can be obtained by setting $k=k_p$ in (\ref{eqn_scd_esti_moving_average1D}). Repeating the averaging process in all virtual pilot positions, we can get $K_f$ accurate CFR estimation samples $\bar{H}_2[k_p],\forall k_p\in\Xi$.

To obtain the $N$-length CFR estimation $\hat H_2$, a Wiener filtering~\cite{Kaiser} based interpolation is performed on these CFR estimation samples:
\begin{eqnarray}
    \label{eqn_1D_wiener}
    \hat H_2[k] = \sum_{k_p\in\Xi}\omega_f[k,k_p]\bar{H}_2[k_p],\quad0\leq k \leq N-1,
\end{eqnarray}
where $\omega_f[k,k_p]$'s are the Wiener filter coefficients computed to minimize the estimation MSE $\mathbb{E}\{\mid H(k) - \hat H_2(k) \mid^2\}$. Rearrange the samples in vector form $\bar{\textbf{H}}_2= \big[\bar{H}_2[k_0],\bar{H}_2[k_1],\ldots,\bar{H}_2[k_{K_f-1}]\big]^T$. Accordingly, the output vector is $\hat{\textbf{H}}_2= \big[\hat{H}_2[0],\hat{H}_2[1],\ldots,\hat{H}_2[N-1]\big]^T$ and the vector form of CFR is $\textbf{H}_2= \big[{H}[0],{H}[1],\ldots,{H}[N-1]\big]^T$. The coefficients of Wiener filter $\omega_f$ are expressed as:
\begin{eqnarray}
    \label{eqn_coef_1D_wiener}
    \boldsymbol \omega_f = \boldsymbol \Phi_f^{-1}\boldsymbol \theta_f,
\end{eqnarray}
where $\boldsymbol \omega_f$ is the $K_f\times N$ coefficient matrix of Wiener filter with the $(m,n)$th element $\omega_f[m,n]$, $\boldsymbol \Phi_f=\mathbb{E}\{\bar{\textbf{H}}_2 \bar{\textbf{H}}_2^H\}= (\textbf{R}^{(1)}_f +\sigma_{\bar{H}_2}^2 \textbf{I})$ is the $K_f\times K_f$ autocorrelation matrix of CFR estimation samples on pilot positions obtained in (\ref{eqn_scd_esti_moving_average1D}), $\sigma_{\bar{H}_2}^2$ is the variance of the estimation error $\bar{H}_2$ derived in (\ref{eqn_mse_1D_averaged}), $\textbf{I}$ is the identity matrix. $\boldsymbol \theta_f = \mathbb{E}\{\bar{\textbf{H}}_2 \textbf{H}\}=\textbf{R}_f^{(2)}$ is the $K_f\times N$ cross-correlation matrix of CFR estimates $\bar{\textbf{H}}_2$ and the real CFR $\textbf{H}$. $\textbf{R}_f^{(1)}$ and $\textbf{R}_f^{(2)}$ are the autocorrelation matrix of the real CFR in the frequency domain, and its components can be computed by (\ref{eqn_corr_freq}). Note that the coefficients can be computed in either non-adaptive~\cite{Kaiser} or adaptive manners~\cite{Sgraja}. The MSE of the CFR estimate after frequency domain filtering is:
\begin{equation}
\label{eqn_mse_1D_wiener}
  \varepsilon_{\hat{H}_2,1D} = \frac{1}{N}\mathrm{Tr}\left(\textbf{R}_f^{(3)}-\textbf{R}_f^{(2)T} \boldsymbol \Phi_f^{-1} \textbf{R}_f^{(2)\ast}\right),
\end{equation}
where $\mathrm{Tr}(\cdot)$ is the trace operation, $\textbf{R}_f^{(3)}$ is the $N\times N$ autocorrelation matrix of CFR.

\begin{figure}[!t]
\centering
\includegraphics[width=3.5in]{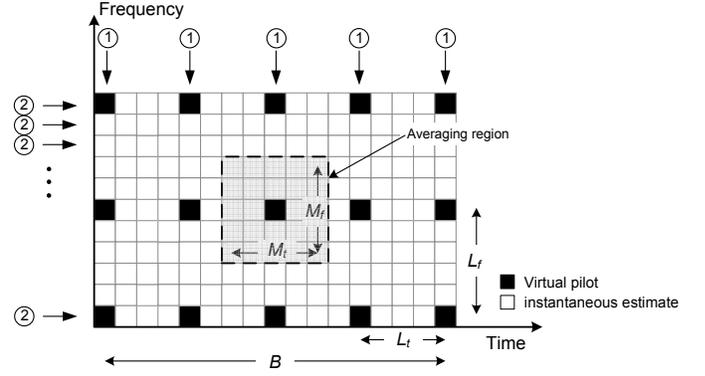}
\caption{2-D averaging and interpolating. The averaging is performed in the 2-D averaging region (shaded area). The interpolation is first carried out in the frequency domain (\textcircled{1}) and consequently in the time domain (\textcircled{2}).}
\label{fig_2D_av}
\end{figure}
\subsection{2-D Refinements}
As presented in above, 1-D refinement approaches only exploit the frequency domain correlation property of the channel. More processing gain can be acquired  if the time domain correlation property is also taken into account.
\subsubsection{2-D Moving Average}
Expanding the average range in (\ref{eqn_scd_esti_moving_average1D}) to 2-D, namely to several consecutive OFDM symbols, more instantaneous channel estimates can be involved in the moving average process. With a proper selection of the averaging length in the time domain with respect to the time variation of the channel, the 2-D moving averaged results are expected to be better than the 1-D counterpart. More specifically, the subcarrier indices within the 2-D averaging region with the $k$\superscript{th} subcarrier of the $i$\superscript{th} OFDM symbol its central position is given by the set $\Theta_{i,k}=\big\{ (p,q)\big| i-\lfloor\frac{M_t-1}{2}\rfloor \leq p \leq i+ \lfloor\frac{M_t-1}{2}\rfloor, k-\lfloor\frac{M_f-1}{2}\rfloor \leq q \leq k+ \lfloor\frac{M_f-1}{2}\rfloor \big\}$. The 2-D moving-averaged CFR estimates are written as:
\begin{eqnarray}
 \label{eqn_scd_esti_moving_average2D}
 && \bar{H}_2[i,k]=\frac{1}{M_tM_f}\sum_{p,q\in\Theta_{i,k}} \tilde{H}_2[p,q] \nonumber\\
 &\approx& H[i,k]+\frac{1}{M_tM_f}\sum_{p,q\in\Theta_{i,k}}\frac{\hat X[p,q]^{\ast}W^{''}}{\eta_{\hat X}[p,q]}.
\end{eqnarray}

The variance of the estimation error in (\ref{eqn_scd_esti_moving_average2D}) is then:
\begin{eqnarray}
\label{eqn_mse_2D_MA}
 &&\sigma_{\bar{H}_2}^2[i,k]=\mathbb{E}\{|H[i,k]- \bar{H}_2[i,k]|^2\}\nonumber\\
 &=&\mathbb{E}\Big\{\Big|\frac{W^{''}}{M_tM_f}\!\sum_{p,q\in\Theta_{i,k}}\!\!\!\frac{\hat X[p,q]}{\eta_{\hat X}[i,k]}\Big|^2\Big\}\nonumber\\
 &=&\frac{\sigma_{W^{''}}^2 }{(M_tM_f)^2}\sum_{p,q\in\Theta_{i,k}}\frac{1}{\eta_{\hat X}[p,q]}.
\end{eqnarray}
The MSE of the 2-D moving averaged CFR is:
\begin{eqnarray}
  &&\varepsilon_{\bar{H}_2,2D}=\frac{1}{N}\sum_{k=0}^{N-1}\mathbb{E} \left\{\big|H[i,k]-\bar{H}_2[i,k]\big|^2\right\} \nonumber\\
 &=&\frac{\sigma_{W^{''}}^2 }{(M_tM_f)^2N}\sum_{k=0}^{N-1}\!\!\sum_{p,q\in\Theta_{i,k}} \!\!\frac{1}{\eta_{\hat X}[p,q]}.
\end{eqnarray}

\subsubsection{2-D Wiener Filtering}
The 2-D estimate refinement and interpolation is carried out over a number of OFDM symbols within an interpolation block as shown in Fig. \ref{fig_2D_av}. The block size $B$, i.e. the number of OFDM symbols in the block, is chosen according to the constraints of latency, storage and complexity.

For low-complexity consideration, some time-frequency indices $(i_p,k_q)\in \Omega$ are pre-selected as virtual pilots. $L_f$ and $L_t$ are the virtual pilot spacings in the frequency and time domains, while $K_f=\lfloor\frac{N}{L_f}\rfloor$ and $K_t=\lfloor\frac{B}{L_t}\rfloor$ are the number of virtual pilots within one OFDM symbol and the number of OFDM symbols including virtual pilots, respectively. The selection of virtual pilots in the frequency domain is discussed in the previous section. The selection in the time domain should also satisfy the sampling theorem. In a similar manner as in the frequency domain, the spacing between virtual symbols in the time domain $L_t$ (in terms of OFDM symbols) is selected so that $L_tT_bf_d\leq 1/4$. Hence, the set of indices of virtual pilots is $\Omega=\big\{(i_p,k_q) \big| i_p=pL_t, 0\leq p\leq K_t-1; k_q=qL_f, 0\leq q \leq  K_f-1 \big\}$.

Then a refined CFR estimate for the virtual pilot located at $(i_p,k_p)$ is computed by averaging all the available CFR estimates within the coherence region $\Theta_{i_p,k_q}$. The averaged estimate can be obtained from (\ref{eqn_scd_esti_moving_average2D}) by setting $i=i_p$ and $k=k_p$. Repeating this process over all virtual pilots in the interpolation block, $K_t\cdot K_f$ refined CFR estimates are acquired. With these more reliable estimates, the overall CFR estimate can be obtained via interpolation.

The 2-D Wiener filtering outperforms other interpolation techniques, e.g. linear interpolation, FFT-based interpolation etc.~\cite{DongXD}. The use of two concatenated 1-D Wiener filters, i.e. one in the frequency domain and the other in the time domain, significantly reduces the computational complexity with negligible performance degradation compared to the optimum 2-D Wiener filter~\cite{Kaiser}. The 2$\times$1-D Wiener filtering is represented as:
\begin{equation}
\label{eqn_2D_wiener_filter}
  \hat H_2[i,k]=\underbrace{\sum_{i_p}\omega_t[i,k,i_p]}_{{\rm{time{\kern 2pt}domain}}}\underbrace{\sum_{k_p}\omega_f[k,i_p,k_p]}_{{\rm{frequency{\kern 2pt}domain}}}\bar{H_2}[i_p,k_p],
\end{equation}
where $\omega_t$'s and $\omega_f$'s are the coefficients of the 1-D Wiener filter in the time and frequency domains, respectively. The coefficients of the frequency domain Wiener filter are the same as those given in (\ref{eqn_coef_1D_wiener}). Let us define the filtered channel estimate matrix $\bar{\textbf{H}}_2$ of size $K_f\times K_t$ whose $(m,n)$th element is $\bar{H_2}[m,n]$. The output of the frequency domain 1-D Wiener filter is denoted as an $N\times K_t$ matrix $\hat{\textbf{H}}_{2}^{f}=\boldsymbol \omega_f^T\bar{\textbf{H}}_2$, where $\boldsymbol \omega_f$ is given in (\ref{eqn_coef_1D_wiener}). The corresponding MSE of the estimation after frequency domain Wiener filtering is shown in (\ref{eqn_mse_1D_wiener}).

The coefficients of the time domain Wiener filter are computed as:
\begin{eqnarray}
   \boldsymbol \omega_t = \boldsymbol \Phi_t^{-1}\boldsymbol \theta_t,
\end{eqnarray}
where
$\boldsymbol \Phi_t=\mathbb{E}\{\hat{\textbf{H}}_2^{fT} \hat{\textbf{H}}_2^{f\ast}\}= (\textbf{R}^{(1)}_t +\sigma_{\hat{H}_2^{f}}^2 \textbf{I})$ is the $K_t\times K_t$ autocorrelation matrix of the filtered CFR estimates $\hat{\textbf{H}}_2^f$, $\sigma_{\hat{H}_2^{f}}^2$ is the variance of the estimation error derived in (\ref{eqn_mse_1D_wiener}). $\boldsymbol \theta_t = \mathbb{E}\{\hat{\textbf{H}}_{2,k}^{fT} \textbf{H}_t^{\ast}\}=\textbf{R}_t^{(2)}$ is the $K_t\times B$ cross-correlation matrix of CFR estimates for the $k$\superscript{th} subcarrier with different time $\hat{\textbf{H}}_{2,k}^{f}$  and the real CFR for a particular subcarrier  with different time $\textbf{H}_t$. Finally, $\textbf{R}_t^{(1)}$ and $\textbf{R}_t^{(2)}$ are the autocorrelation matrix of the real CFR in the time domain, and its components are computed by (\ref{eqn_corr_time}).

The MSE of the CFR estimate after time domain filtering is:
\begin{equation}
\label{eqn_mse_2D_wiener}
  \varepsilon_{\hat H_2,2D} = \frac{1}{B}\mathrm{Tr}\left(\textbf{R}_t^{(3)}-\textbf{R}_t^{(2)T} \boldsymbol \Phi_t^{-1} \textbf{R}_t^{(2)\ast}\right).
\end{equation}

According to the Jakes' model, the coefficients of the time domain Wiener filter are only determined by the Doppler frequency. Note that, since the time domain interpolation length is not very large, the computation of the coefficients of time domain Wiener filter is not prohibitive. It should be also noted that the same coefficients are used for all subcarriers, which indicates that the storage spent for these coefficients is negligible compared with the size of interleaver.

Concerning the frequency domain Wiener filter, the coefficients can be pre-computed according to the uniform delay power spectrum assumption which represents the ``worst case'' of the mobile channel~\cite{Kaiser}. Given a virtual pilot pattern, the values of the coefficients only depend on the maximum delay spread of the channel which can be measured in reality. Therefore, the coefficients can be pre-computed with some typical channel situations. Even more complexity reduction can be achieved by limiting the Wiener filtering on several neighboring pilots~\cite{DongXD}.

\subsection{MMSE Combination}

\begin{table*}[tb]
 \centering
\begin{threeparttable}
 \caption{Complexity and Storage Comparison}
 \label{tbl_cplt_cmp}
 {\renewcommand\arraystretch{1.3}
 \begin{tabular}{|m{1.2cm}| m{2cm}||m{5.5cm}|m{5cm}|}\hline 
 \multicolumn{2}{|c||}{{\bfseries Method}}& {\bfseries Complexity} & {\bfseries Storage}  \\ \hline\hline
 \multicolumn{2}{|c||}{{PN based method}}    & {$\mathcal{O}$($\nu\cdot\log \nu$)} & {$\nu$ complex symbols}  \\ \hline\hline
    \multirow{2}{1.2cm}{Proposed method}   &   Moving average             & $\mathcal{O}$($N$) & \multirow{2}{5cm}{One OFDM symbol for 1-D methods, $B$ OFDM symbols for 2-D methods.} \\ \cline{2-3}
                                            &  Wiener filtering                 & $\mathcal{O}$($N$) to $\mathcal{O}$($N^2$)& \\ \hline\hline
      \multicolumn{2}{|c||}{\textsf{MUCK03}~\cite{PRPOFDM} }   & $\mathcal{O}(\nu\cdot\log\nu)$ when $M\ll\nu$, $\mathcal{O}$($\nu^2$) otherwise & $M$ OFDM symbols~\tnote{a} \\ \hline
\multicolumn{2}{|c||}{{\textsf{YANG08}}~\cite{YANG08TCE} }   & {$\mathcal{O}(\mathcal{K}\cdot\log\mathcal{K})~\tnote{b}$} & {One OFDM symbol and one GI}\\ \hline
    \multicolumn{2}{|c||}{\textsf{WANG05}~\cite{WANG05}}    & $\mathcal{O}$($N\cdot\log N$) & Two OFDM symbols  \\ \hline
    \multicolumn{2}{|c||}{\textsf{TANG07}~\cite{TANG07} }   & $\mathcal{O}((N+\nu)\cdot\log(N+\nu))$ & One OFDM symbol and one GI\\ \hline
    \multicolumn{2}{|c||}{\textsf{STEENDAM07}~\cite{KSPOFDM}}& $\mathcal{O}$($N^3$)& One OFDM symbol\\ \hline
     \multicolumn{2}{|c||}{\textsf{ZHAO08}~\cite{ZHAO08} }   & ($\mathcal{O}$($N$) + complexity of decoding ) for iteration, $\mathcal{O}$($N^2$) or $\mathcal{O}$($N^3$) for final stage & Storage should be greater than the interleaving depth.~\tnote{c}\\ \hline
 \end{tabular}}
 {\footnotesize
\begin{tablenotes}
\item [a] The parameter $M$ is the length of the averaging window. It is set to $20$ to $40$ OFDM symbols for BPSK and QPSK, $40$ to $72$ OFDM symbols for 16QAM and $120$ to $240$ for 64QAM.
\item [b] $\mathcal{K}$ is the size of FFT/IFFT used to build the received PN sequence and make channel estimation. It can be selected from $\nu+L$ to $N$ or even larger. $\mathcal{K}$ is set to 2048 in~\cite{YANG08TCE}.
\item [c] As far as the DTMB system is concerned, the interleaving depth is 170 or 510 OFDM symbols.
\end{tablenotes}
}
\end{threeparttable}
\end{table*}

When both PN-based and data-aided channel estimates are obtained, a linear combination is proposed to get a final CFR estimate:
\begin{equation}
\label{eqn_H_combined}
\hat H = \beta \hat H_1 + (1-\beta)\hat H_2,
\end{equation}
where $\hat H_2$ is a generic expression of the data-aided channel estimations which can be obtained from (\ref{eqn_scd_esti_moving_average1D}), (\ref{eqn_1D_wiener}), (\ref{eqn_scd_esti_moving_average2D}) or (\ref{eqn_2D_wiener_filter}) using different techniques.
The combination is carried out only for the active subcarriers while the CFR estimates for null subcarriers are kept as the results obtained from the PN-based one. The optimum weight values $\beta_{opt}$ can be obtained using the MMSE criteria:
\begin{eqnarray}
  &&\beta_{opt}=\arg \min_\beta \left\{ \mathbb{E}\{| H - \hat H|^2\}\right\} \nonumber\\
  &=&\arg \min_\beta \left\{\beta^2 \varepsilon_{\hat H_1}+(1-\beta)^2 \varepsilon_{\hat H_2}\right\}.
\end{eqnarray}
The above equation uses the fact that $\hat H_1$ and $\hat H_2$ are obtained from different sources and thus uncorrelated. Since the MSE function is convex, the $\beta_{opt}$ is obtained by setting the derivative with respect to $\beta$ equal to zero and the solution is:
\begin{equation}
\beta_{opt} = \frac{\varepsilon_{\hat H_2}}{\varepsilon_{\hat H_1}+\varepsilon_{\hat H_2}},
\end{equation}
where $\varepsilon_{\hat H_1}$ is MSE of the PN-based estimation obtained from (\ref{eqn_mse_pn_est}), while $\varepsilon_{\hat H_2}$ is the MSE of the data-aided channel estimation and is computed from (\ref{eqn_mse_1D_averaged}), (\ref{eqn_mse_1D_wiener}), (\ref{eqn_mse_2D_MA}) or (\ref{eqn_mse_2D_wiener}) depending on different methods.

The proposed method can be carried out in an iterative manner. More precisely, if it has not achieved the preset maximum iteration times yet, the combined CFR estimates (\ref{eqn_H_combined}) are used as the initial channel estimates of the next channel estimation iteration for PN subtraction and symbol equalization. The equalized data symbols using the updated channel estimates are more accurate than previous iterations. Hence, more reliable data-aided channel estimates can be acquired in the new iteration. If the the maximum iteration times are achieved, the combined channel estimates output as the final channel estimates for the data equalization and FEC decoding.

\section{Complexity Analysis}

In this section, we analyze the additional computational complexity introduced by the proposed data-aided channel estimation method.
The complexity is evaluated in terms of required real  multiplications and real additions for each OFDM symbol per iteration. In this paper one complex multiplication is counted as $2$ real additions and $4$ real multiplications although there exist some smarter ways requiring fewer multiplication times. The data rebuilding process including (\ref{eqn_prb_bits}), (\ref{eqn_prb_constellation}) and (\ref{eqn_data_rebuild}) needs $(N\cdot\mu\log_2\mu+N\cdot2\log_2\mu)$ multiplications and $N\cdot 2\log_2\mu$ additions. The instantaneous data-aided estimation (\ref{eqn_scd_estimation}) needs $8N$ multiplications and $3N$ additions.

The computational complexities required by the estimate refinement vary with respect to different strategies. For the 1-D moving average method, the averaging for each subcarrier with an averaging length of $M$ requires $2(M-1)$ additions and $2$ multiplications neglecting the edge effect. Hence, the moving averaging over $N$ subcarriers requires $2(M-1)N$  additions and $2N$ multiplications. The combination of the two channel estimates requires $2N$ additions and $4N$  multiplications. The overall basic operations are $\big((\mu+2)\log_2\mu+14\big)N$ multiplications and $(2\log_2\mu+2M+3)N$  additions for each iteration. Given $M\ll N$, the additional complexity of the 1-D moving average based channel estimation is $\mathcal{O}(N)$.

The 2-D moving average method, on the other hand, has generally the same computational complexity except that the averaging process for each subcarrier requires $2(M_tM_f-1)N$ additions and $2N$ multiplications and the overall required additions turns to $(2\log_2\mu+2M_tM_f+3)N$. Given the condition $M_tM_f\ll N$ fulfilled, the complexity is still $\mathcal{O}(N)$.

As far as the Wiener filtering based approaches are concerned, for the $K$ virtual pilot in each OFDM symbol, the averaging of the instantaneous estimates in the coherence bandwidth requires $2(M-1)K$ real additions and $2K$ real multiplications. As the coefficients of the Wiener filter is real~\cite{Kaiser}, the Wiener filtering needs $2KN$ multiplications and $2KN$ additions. Taking the symbol rebuilding and combination into account, the overall operations for 1-D Wiener filtering based method is $(2\log_2\mu+5+2K)N+2(M-1)K$ additions and $\big((\mu+2)\log_2\mu+12+2K\big)N+2K$ multiplications, which is between $\mathcal{O}(N)$ and $\mathcal{O}(N^2)$.

For the 2-D Wiener filtering case, in each block consisting of $B$ OFDM symbols, the averaging and 1-D frequency domain Wiener filtering is repeated $K_t$ times which counts $2(N+M_tM_f-1)K_fK_t$ additions and $2(N+1)K_fK_t$ multiplications. Then, the time domain Wiener filtering needs $2K_tBN$ multiplications and additions. Therefore, the overall operations per OFDM symbol per iteration for 2-D Wiener filtering based method is $\big((\mu+2)\log_2\mu+12+2\frac{K_f}{L_t}+2K_t\big)N+\frac{2K_f}{L_t}$ multiplications and $(2\log_2\mu+5+2\frac{K_f}{Lt}+2K_t)N+2\frac{(M_tM_f-1)K_tK_f}{B}$ additions, which is, similar to the 1-D Wiener filtering case, between $\mathcal{O}(N)$ and $\mathcal{O}(N^2)$. Even lower complexity can be achieved by reducing the size of the Wiener filter.

\begin{figure}[!t]
\centering
\includegraphics[width=3.5in]{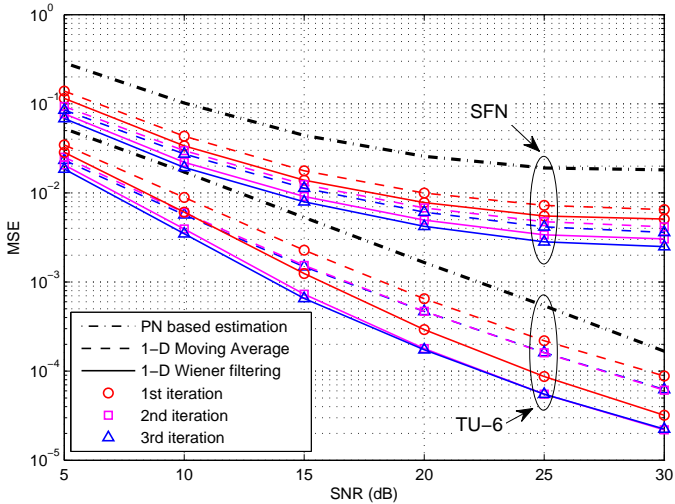}
\caption{MSE of the proposed data-aided channel estimation using 1-D Moving Average and 1-D Wiener filtering with QPSK, in the TU-6 channel and SFN channels with velocity of 30 km/h.}
\label{fig_MSE_1D_MA_WF_TU6_SFN}
\end{figure}

The computational complexities of proposed method and several typical existing methods~\cite{KSPOFDM,PRPOFDM,WANG06,TANG07,YANG08TCE,ZHAO08}  are shown in Table~\ref{tbl_cplt_cmp}. From the comparison, we find that the method {\small\textsf{MUCK03}} achieves least computational complexity but requires a large amount of storage. The proposed method using moving average needs less complexity than rest of other methods. In the meantime, the complexity and storage costs of proposed method using Wiener filtering are close to the methods {\small\textsf{WANG05}}, {\small\textsf{TANG07}} and {\small\textsf{YANG08}}. The methods {\small\textsf{STEENDAM07}} and {\small\textsf{ZHAO08}} spend more complexity than other techniques.

\begin{figure}[!t]
\centering
\includegraphics[width=3.5in]{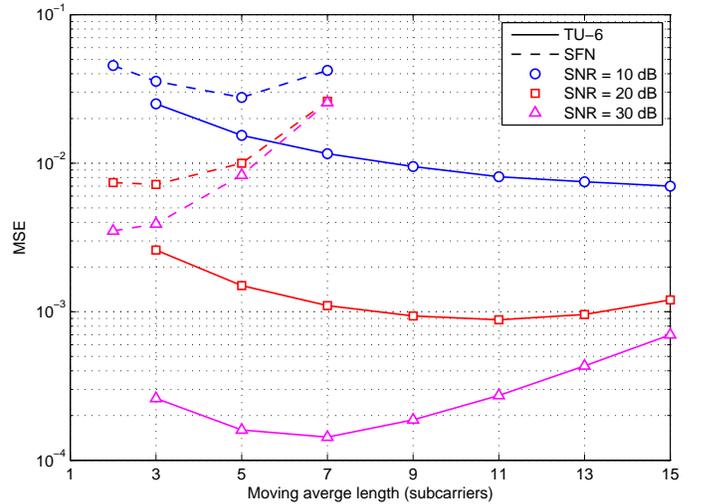}
\caption{MSE of the 1-D Moving average method \textit{before} combination with different averaging lengths.}
\label{fig_MSE_different_MA_Length}
\end{figure}

\section{Simulation Results}
\subsection{Simulation Settings}
Simulation parameters are chosen from the DTMB system~\cite{DTMB}. Baseband signal bandwidth is $7.56$ MHz. FFT size is $3780$ which results in a subcarrier spacing of $2$ kHz. All subcarriers are active. The length of the GI is set to $\nu=420$. The power of the PN sequence is twice as much as data symbols. The COST207 Typical Urban (TU-6) channel model~\cite{COST207} is employed in the simulation. The proposed algorithm is also evaluated in the single frequency network (SFN) scenario which is a spectrum efficient solution widely used in broadcast networks. In this case, the same signal is sent from two different transmitters at the same time on the same carrier frequency. The signal from these two transmitters experiences independent fading. The equivalent CIR of the SFN channel is the combination of the CIR's of two independent TU-6 channels. The propagation distance difference between the two signals causes time delay and power attenuation on the second CIR~\cite{LIUICC}. In this paper, the distance difference is set to $7$ km corresponding to  $23.33$ $\mu s$ time delay which makes the overall length of the CIR longer than that of the CP of the PN. The power attenuation is set to $10$ dB. Following the computation in \cite{Rappaport}, when the frequency correlation function is $0.9$, the coherence bandwidths of the TU-6 channel and the SFN channel are $18.8$ kHz and $2.94$ kHz, respectively. Given the $2$ kHz subcarrier spacing in the DTMB system, the frequency domain averaging length $M$ is accordingly set to $9$ and $3$ for the TU-6 and SFN channels, respectively. The virtual pilot spacings in the frequency and time domains are set equal to the averaging lengths in each domain, respectively.

\subsection{MSE Performance}
\begin{figure}[!t]
\centering
\includegraphics[width=3.5in]{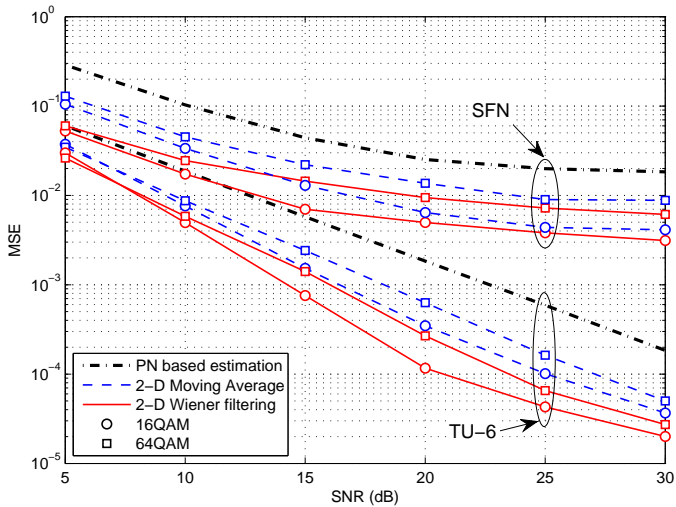}
\caption{MSE performance of the proposed method after two iterations using 2-D moving average and Wiener filtering methods. $M_t=2$, $M_f=9$ for the TU-6 channel and $M_t=2$, $M_f=3$ for the SFN channel. Velocity is set to 6km/h for both TU-6 and SFN channels.}
\label{fig_MSE_2D_WF_MA}
\end{figure}

Fig.\ref{fig_MSE_1D_MA_WF_TU6_SFN} presents the MSE performance of the proposed algorithm with QPSK, using 1-D moving average and 1-D Wiener filtering in the TU-6 and SFN channels, respectively. It can be observed that the accuracy of the channel estimation is progressively improved. The iterative estimation process converges very fast. A significant improvement can be obtained after only two iterations. More specifically, in the TU-6 channel, when the 1-D moving average is used, the proposed data-aided channel estimation method acquires about $4.1$ dB gain in terms of required SNR to achieve an MSE level of $1\times10^{-2}$ compared with the PN-based one. If the more powerful 1-D Wiener filtering technique is adopted, the gain increases to $5.1$ dB. The higher the SNR, the more efficient the Wiener filtering. Moreover, it should be noted that the performance of using Wiener filtering after one iteration is as good as the best performance that using moving average technique can achieve. This provides a less processing delay (less iterations) trade-off with higher computational complexity for each iteration. On the other hand, in the SFN channel, since the delay spread is much longer than the CP of the PN sequence, there is a strong ISI on the PN sequence. This can be observed from the fact that the performance of the PN-based channel estimation is seriously degraded and appears an estimation error floor at high SNR. Furthermore, as the averaging length is much shorter in the SFN case, the noise mitigation ability is limited in the data-aided channel estimation. Even though in such a harsh channel, the data-aided method however offers $6.9$ dB and $8.1$ dB gain in terms of required SNR to achieve MSE of $5\times10^{-2}$ compared to the PN based method, when using the 1-D moving average and Wiener filtering, respectively. Moreover, the estimation performance is improved approximately ten times at high SNR region.

Fig.\ref{fig_MSE_different_MA_Length} depicts the impact of the averaging length to the performance of 1-D moving average based method. At a low SNR, e.g. $10$ dB, the MSE is monotonically decreasing with the increase of the averaging length. However, at a higher SNR, e.g. at $30$ dB, the MSE of the averaged estimation results is degraded with a long averaging length. Hence, a moderate averaging length, say $M=7\sim9$, is a proper choice. Considering that the data-aided channel estimation is more crucial in lower SNR region, it is better to bias our selection to a greater length. Therefore, $M=9$ is a good trade-off that suits all noise level for the TU-6 channel, which coincides with the selection according to the coherence bandwidth. Similar conclusion can be drawn in the SFN channel. $M=3$ is the best trade-off of the averaging length in the SFN channel.

\begin{figure}[!t] 
\centering
\includegraphics[width=3.5in]{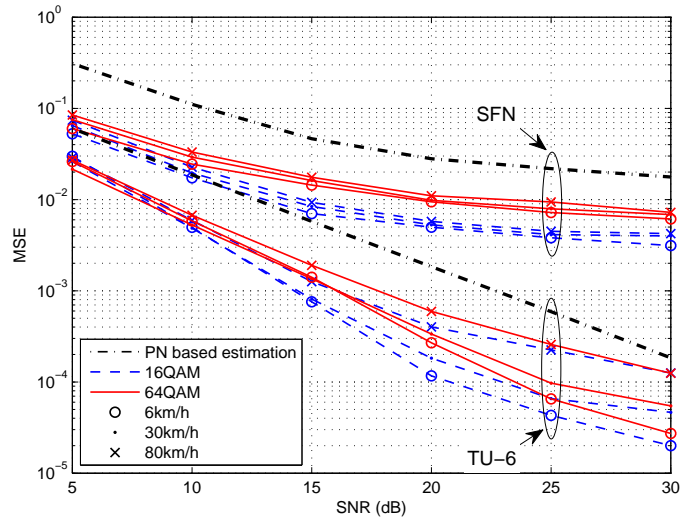}
\caption{MSE performance of the proposed method using 2-D Wiener filtering method after two iterations with different velocities.  $M_t=2$, $M_f=9$ for the TU-6 channel and $M_t=2$, $M_f=3$ for the SFN channel.}
\label{fig_MSE_different_speeds}
\end{figure}

Fig.\ref{fig_MSE_2D_WF_MA} shows the MSE of proposed 2-D method after two iterations. As the 1-D method is proved to be effective with QPSK, the 2-D method is only evaluated with higher order constellations, namely 16QAM and 64QAM. The averaging length in the time domain is selected to $M_t=2$ which is a very conservative setting that only requires the channel keeping similar within two consecutive OFDM symbols. Even with such a short averaging length in the time domain, the 2-D method can still provide effective estimation. For instance, the proposed method using 2-D Wiener filtering acquires $8.4$ dB gain over the PN-based method in terms of the required SNR to achieve an MSE level of $1\times 10^{-3}$ with 16QAM in the TU-6 channel. In the SFN channel, the improvement is also significant. The estimation error floor is reduced from $2\times 10^{-2}$ to $3\times 10^{-3}$ with 16QAM at a SNR of $30$ dB. Given longer time domain averaging length, more gain can be expected in a static channel.

Fig.\ref{fig_MSE_different_speeds} presents the changing of the MSE performance with different velocities of the receiver in the TU-6 and SFN channels. With the increase of the velocity, the variation of the channel among consecutive OFDM symbols is more and more notable, which limits the accuracy in the averaging results in (\ref{eqn_scd_esti_moving_average2D}). In low SNR region (e.g. SNR less than $10$ dB), the channel variation is a less significant influence that affects the estimation performance compared to the noise. Whereas in a higher SNR region where the noise is no longer the dominant factor, the channel variation compromises the performance of the proposed method. However, it should be noted that the proposed method can still provide adequate improvement. For instance, in the TU-6 channel with a velocity of $30$ km/h, the proposed method obtains $8.3$ dB gain with 16QAM in terms of required SNR to achieve MSE of $1\times10^{-3}$. While in SFN channel, as the ISI is the dominant factor that affects the MSE results. Therefore, the impact of the channel variation is not significant.

\begin{figure}[!t] 
\centering
\includegraphics[width=3.5in]{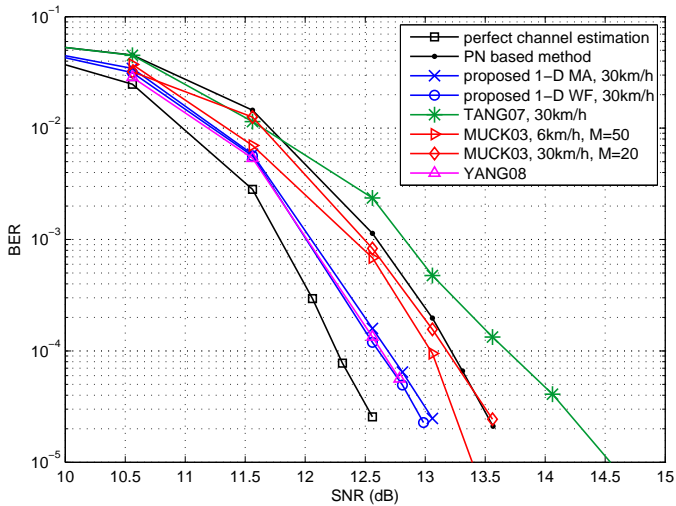}
\caption{BER of the DTMB system with QPSK and LDPC (R=0.8) using different channel estimation methods in the TU-6 channel with velocity of 30 km/h. Iteration time is set to 2 for the proposed data-aided channel estimation in both 1-D Moving Average (MA) and 1-D Wiener Filtering (WF) cases.}
\label{fig_BER_4QAM_TU6}
\end{figure}

\begin{figure}[!t] 
\centering
\includegraphics[width=3.5in]{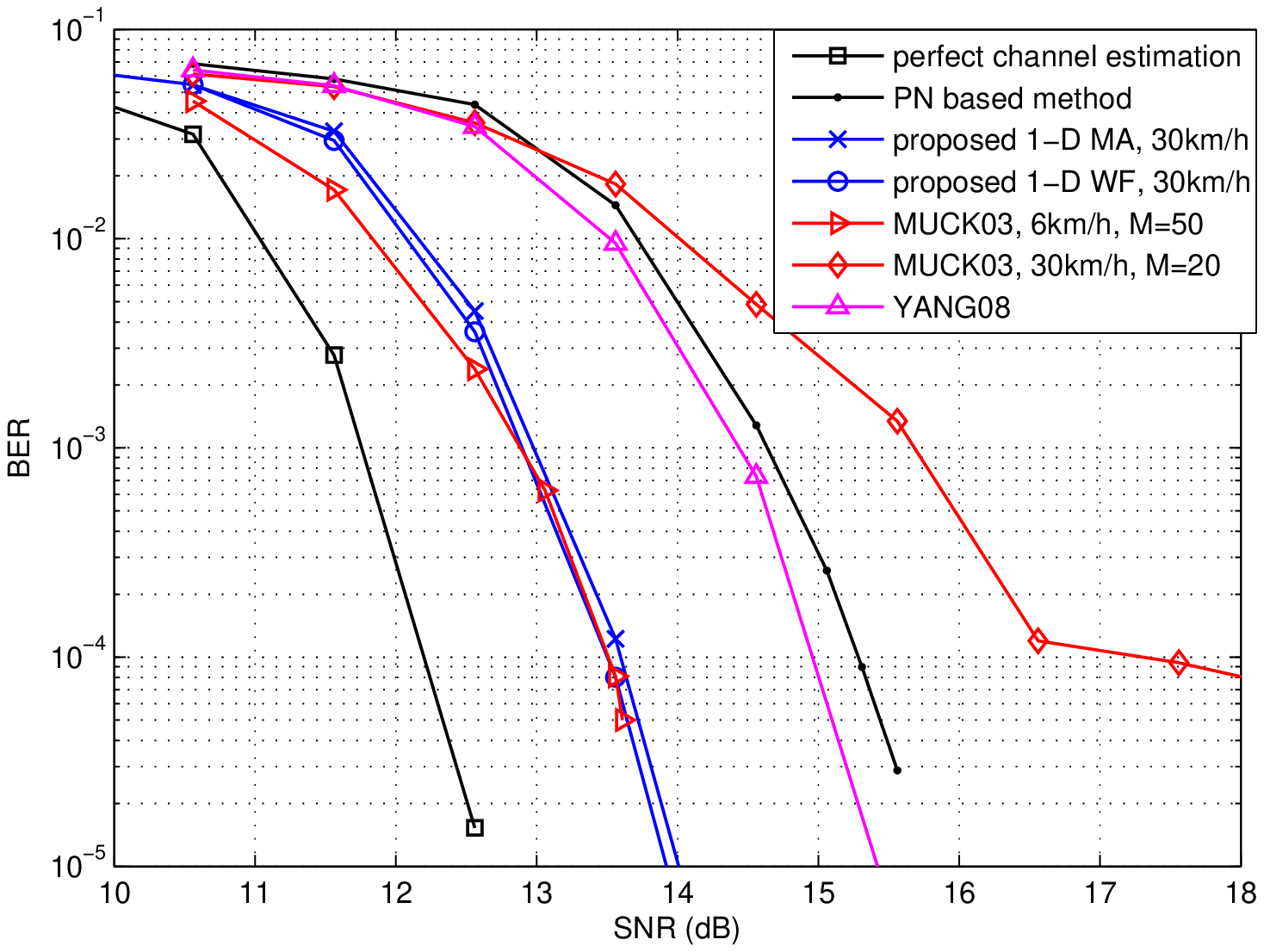}
\caption{BER of the DTMB system with QPSK and LDPC (R=0.8) using different channel estimation methods in the SFN channel with velocity of 30 km/h. Iteration time is set to 2 for the proposed data-aided channel estimation in both 1-D Moving Average (MA) and 1-D Wiener Filtering (WF) cases.}
\label{fig_BER_4QAM_SFN}
\end{figure}

\begin{figure}[!t] 
\centering
\includegraphics[width=3.5in]{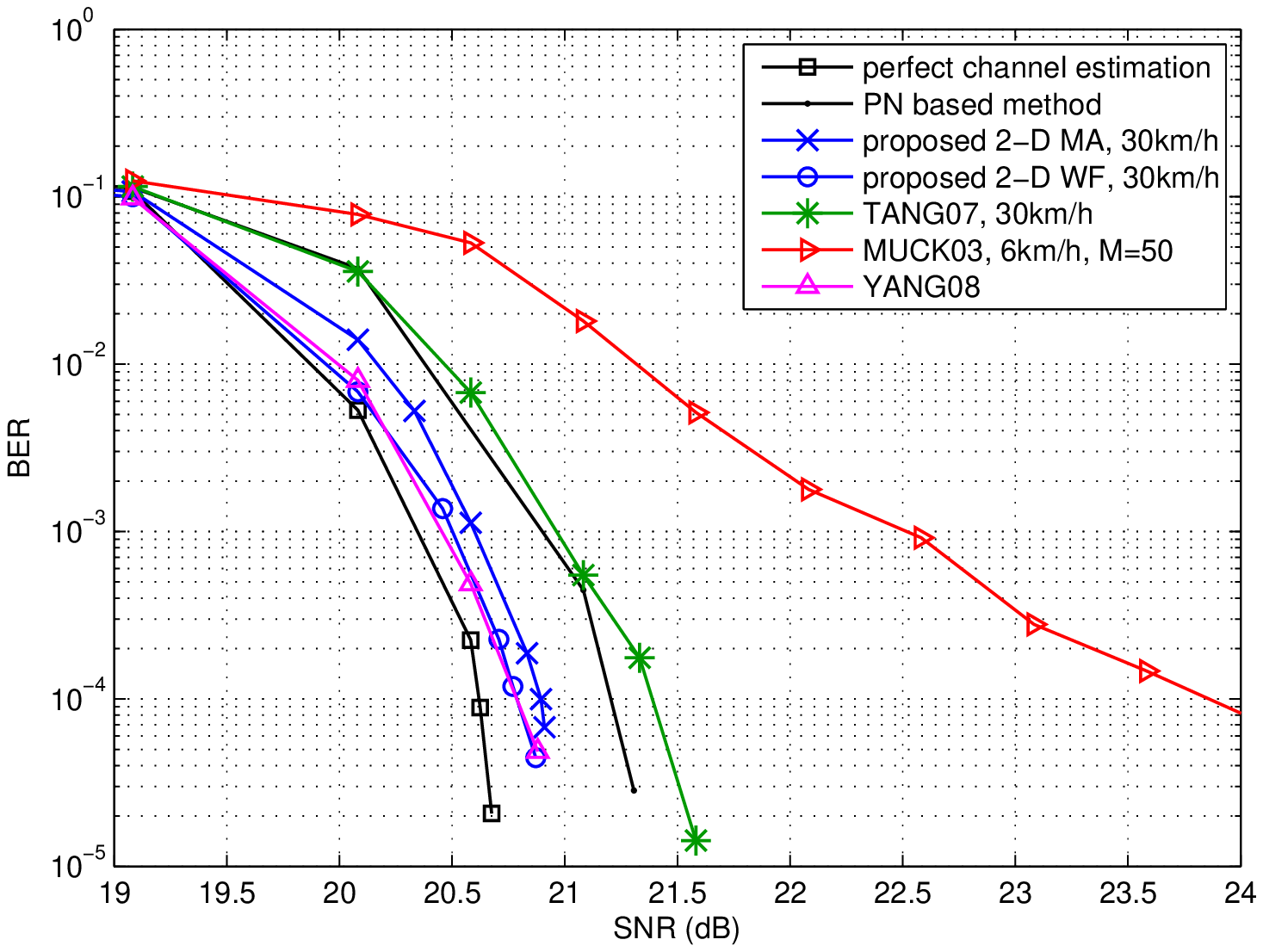}
\caption{BER of DTMB system with 64QAM and LDPC (R=0.6) using different channel estimation methods in the TU-6 channel with velocity of 30km/h. $M_t=2$,  $M_f=9$ for Moving Average (MA) and Wiener Filtering (WF) cases.}
\label{fig_BER_64QAM_TU6}
\end{figure}

\begin{figure}[!t] 
\centering
\includegraphics[width=3.5in]{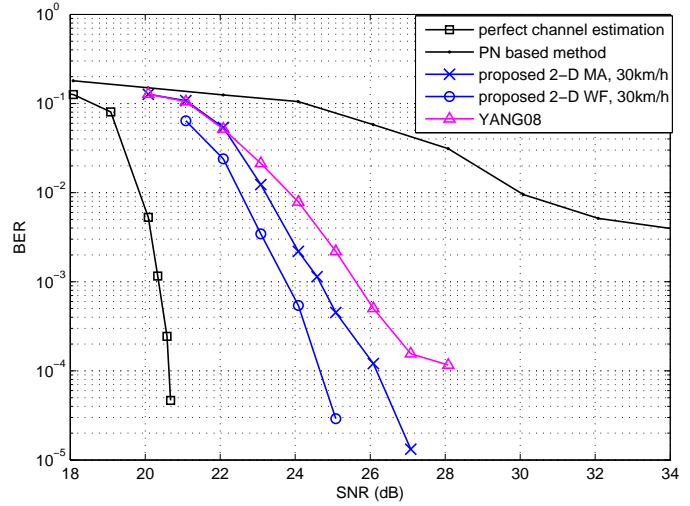}
\caption{BER of DTMB system with 64QAM and LDPC (R=$0.6$) using different channel estimation methods in the SFN channel with velocity of $30$ km/h. $M_t=2, \ M_f=3$ for both moving average and Wiener filtering cases.}
\label{fig_BER_64QAM_SFN}
\end{figure}

\subsection{BER Performance}
In this subsection, we present the bit error rate (BER) performance of the DTMB system using the proposed data-aided channel estimation method. Both the convolutional interleaver, BCH code and LDPC code are included in the simulation in order to give a system level evaluation. The convolutional interleaver is set to $(52, 240)$ which corresponds to a time delay of $170$ OFDM symbols. The mobile speed is set to $30$ km/h which introduces a maximum Doppler frequency of $13.89$ Hz, given the carrier frequency $500$ MHz. The coherence time is accordingly $0.0305$ s and equivalent to the duration of 61 OFDM symbols (without GI). The proposed method is compared with the PN-based method as well as three important methods in the literatures, namely {\small\textsf{MUCK03}}, {\small\textsf{TANG07}} and {\small\textsf{YANG08}}. Since the {\small\textsf{MUCK03}} method is the least complex one in the literature and the  {\small\textsf{TANG07}} and {\small\textsf{YANG08}} methods have comparative complexity as the proposed one while providing superior performance than its predecessors such as {\small\textsf{WANG05}}, it is interesting to take them as reference methods.

\begin{table*}[t]
\centering
\begin{threeparttable}
\caption{Summary of the gains obtained using the data-aided channel estimation methods.}
\label{tbl_Gains}
\footnotesize
\renewcommand\arraystretch{1.3}
\begin{tabular}{|m{0.8cm}|m{1.2cm}|m{1.9cm}||m{1.5cm}|m{1.5cm}| m{1.7cm}|m{1.5cm}||m{1.8cm}|}
\hline
 {Channel} & {Modulation} & {Method} & Gain over PN-base method~(dB) & Gain over method {\scriptsize\textsf{TANG07}}~(dB)& Gain over method {\scriptsize\textsf{MUCK03}}~(dB)\tnote{a}& Gain over method {\scriptsize\textsf{YANG08}}~(dB) & Gap~away~from perfect~estimation~(dB) \\
\hline\hline
\multirow{4}{0.8cm}{{TU-6}} & \multirow{2}{1.2cm}{QPSK} & Moving average & 0.5 & 1.1 & 0.5~[0.3] & -0.05 & 0.5 \\ \cline{3-8}
 & & Wiener filtering & 0.6 & 1.2 & 0.6~[0.4]  & 0  & 0.4 \\ \cline{2-8}
 & \multirow{2}{1.2cm}{64QAM} & Moving average & 0.3 & 0.5 & --~[$>3.0$]  & -0.05 & 0.3 \\ \cline{3-8}
 & & Wiener filtering & 0.4 & 0.6 & --~[$>3.1$]  & 0 & 0.2 \\
\hline\hline
\multirow{4}{0.8cm}{{SFN}}&  \multirow{2}{1.2cm}{QPSK} & Moving average & 1.7 & -- & $>4.3$~[-0.1]  & 1.38 & 1.3\\
\cline{3-8}
 & & Wiener filtering & 1.8 & -- & $>4.4$~[0.04]  & 1.46 & 1.2  \\ \cline{2-8}
 &  \multirow{2}{1.2cm}{64QAM} & Moving average & $>7.0$ & -- & --  & $>2.0$ &  5.8\\ \cline{3-8}
 & & Wiener filtering & $>8.0$ & -- & -- & $>3.0$ &   4.2  \\
\hline
\end{tabular}
{\footnotesize
\begin{tablenotes}
\item [a] The first value is obtained with $M=20$ given a velocity of $30$ km/h, while the value in the square brackets is obtained with $M=50$ given a velocity of $6$ km/h.
\end{tablenotes}
}
\end{threeparttable}
\end{table*}

Fig.\ref{fig_BER_4QAM_TU6} and Fig.\ref{fig_BER_4QAM_SFN} present the BER performance with QPSK in the TU-6 and SFN channels, respectively, while Fig.\ref{fig_BER_64QAM_TU6} and Fig.\ref{fig_BER_64QAM_SFN} depict the BER with the 64QAM in these two channels. Note that the method {\small\textsf{TANG07}} does not adapt to the SFN channel. Therefore, its performance is not shown in that case. As the method {\small\textsf{MUCK03}} is affected by the channel variations, its performance is presented with low mobility ($6$ km/h) and medium mobility ($30$ km/h) cases. To quantify the performance of each method, we observe the required SNR to achieve a BER level of $5\times10^{-5}$. The gains obtained by the proposed method are listed in Table~\ref{tbl_Gains}. There are some interesting observations to be highlighted:

1. \textit{The proposed method outperforms the reference methods.}
The proposed method achieves better BER performance than the reference methods in almost all cases considered in the simulations. More specifically, it acquires more than $0.5$ dB, $0.3$ dB and $1.1$ dB gains over the PN-based, {\small\textsf{MUCK03}} and {\small\textsf{TANG07}} methods, respectively, with QPSK in the TU-6 channel. When used with 64QAM, the improvements are accordingly more than $0.3$ dB, $3$ dB and $0.5$ dB, respectively. In the more harsh SFN channel case, the improvements are more obvious. The proposed method provides $1.7$ dB and $4.3$ dB gains over the PN-based method and {\small\textsf{MUCK03}} method with QPSK, respectively. Compared with {\small\textsf{YANG08}} method, the proposed method achieves similar performance in the TU-6 channel. Yet, we also notice that after the partial-decision-aided interference removal there still exists residual interference in the PN sequence. The impact of the residual interference is not serious when the channel delay spread is short (for instance in the TU-6 channel). However, it becomes more harmful in the SFN channel. The performance of {\small\textsf{YANG08}} method significantly degrades in SFN cases. The proposed method outperforms {\small\textsf{YANG08}} method more than $1.3$ dB and $2.0$ dB with QPSK and 64QAM, respectively. In addition, the residual gap between the proposed method and the perfect channel estimation case turns out to be very small. Especially in the TU-6 channel, the proposed method is only $0.2$ dB to $0.5$ dB away from the case using perfect channel estimation.

2. \textit{The proposed method is robust in different channel conditions.} It can steadily provide satisfactory performance in time-varying channels as well as the channels with long time delay spread. The method {\small\textsf{MUCK03}} eliminates the ISI by averaging the known sequence over a large number of OFDM symbols and thus requires the channel to remain unchanged within the averaging window. It is consequently not robust in face of the time-varying channel condition. In addition, it is not efficient  with higher order constellations such as 64QAM. In contrast, the proposed method performs channel estimation over only either one (1-D case) or two OFDM symbols (2-D case). Therefore, it has stronger immunity to the channel variations. Moreover, the SFN channel has extremely long time delay spread with strong echoes and is thus recognized as the most difficult situation for the channel estimation. For instance, the method {\small\textsf{TANG07}} suffers from the interference from previous OFDM symbols and does not adapt to the SFN channel. The performance of {\small\textsf{YANG08}} method significantly degrades in the SFN channel due to the interference as well. In contrast, the proposed method exploits the OFDM data symbols to refine the channel estimates obtained from the PN-based estimation, which can provide reliable channel estimation in presence of long ISI.

3. \textit{Both the moving average and the Wiener filtering techniques provide satisfactory performance with different channel conditions.}
The performance difference between two methods is small in the TU-6 channel and in the SFN channel with QPSK. That is to say it is convenient to use the low-complexity moving average based estimation in the common case. However, the Wiener filtering technique performs better with 64QAM in the harsh SFN channel where it requires more reliable channel estimation. Moreover, it has been shown in the MSE simulation that the Wiener filtering technique needs fewer iterations to achieve a satisfactory estimation results. In other words, it is suitable for the application that has higher restriction of processing time delay.

Note that all the results are obtained using the powerful LDPC code. More gains can be expected when weaker channel codings are used.

\section{Conclusion}
This paper proposes a novel efficient solution to the challenging channel estimation task in the TDS-OFDM system. The presented algorithm takes basis on the classical PN-based estimation and processes estimation refinement by exploiting a low-complex data-aided approach. In contrast to a classical turbo channel estimation case, this new algorithm does not include the decoding and interleaving functions when rebuilding the data symbols in order to reduce the computational complexity. This enables the new proposed method to be integrated in systems which contain extremely long interleaver and sophisticated channel decoder like the DTMB system. The rebuilt soft data symbols are used as virtual pilot for data-aided channel estimation. In order to suppress the noise existing in the data-aided channel estimates, several techniques including moving average and Wiener filtering in 1-D and 2-D have been conducted. The cooperation of these refining techniques and the soft symbol feedback enables reliable data-aided channel estimation without intensive computation. Simulation results show that the new proposed data-aided channel estimation algorithm provides satisfactory performance even often very close to the perfect channel estimation case. Compared to already existing techniques, the proposed method performs very well in various transmission cases, including the 64QAM constellation, long delay spread channels and mobility scenarios. Note that the use of the novel low-complex data-aided channel estimation algorithm presented in this paper is not limited in TDS-OFDM. In fact, it can be easily adopted by the traditional CP-OFDM systems where the pilot assisted channel estimate will serve as the initial estimate.


\appendix[Interference Resulted from Imperfect PN Removal]
Take (\ref{eqn_received_sig_after_pn_rml}) and (\ref{eqn_ola}), and suppose that the PN sequences are identical for all OFDM frames, the residual PN sequence in the received signal is written as:
\begin{eqnarray}
\xi[n] 
  &=& \sum_{l=0}^{L-1} \Delta h_lc[n-l]_{\nu},\quad 0\leq n<\nu.
\end{eqnarray}
The interference for the  $k$\superscript{th} subcarrier is thus:
\begin{eqnarray}
  &&I[k] = \frac{1}{\sqrt{N}}\sum_{n=0}^{N-1}\xi[n]e^{-j\frac{2\pi}{N}nk} \nonumber\\
  &=& \frac{1}{\sqrt{N}} \sum_{l=0}^{L-1} \Delta h_l \sum_{n=0}^{\nu-1} c[n-l]_{\nu}e^{-j\frac{2\pi}{N}nk}.
\end{eqnarray}
Denote $c[n-l]_{\nu}$ by $c_l[n]$, representing a circular shift of $l$ bits of sequence $c[n]$ towards right side.

Using the US assumption, i.e. $\mathbb{E}[h^{\ast}_{l}h^{\ }_{l'}]=\sigma_{l}^2\delta[l-l']$, the power of the interference is:
\begin{eqnarray}
&&\sigma_I^2[k]=\mathbb{E}\big\{I[k]^{\ast}I[k]\big\}\nonumber\\
&=&\frac{1}{N}\!\!\sum_{l=0}^{L-1}\sigma_{\Delta h_l}^2 \!\!\!\sum_{n_1, n_2=0}^{\nu-1}\!\!\! c_l[n_1]^{\ast}c_l^{\ }[n_2]e^{-j\frac{2\pi}{N}(n_2-n_1)k},\nonumber
\end{eqnarray}
where $\sigma_{\Delta h_l}^2=\mathbb{E}[|\Delta h^{\ }_{l}|^2]$ is variance of the estimation error of the  $l$\superscript{th} channel tap. Substituting $n_1$, $n_2$ by $q=n_2-n_1$ and $n=n_1$, it yields:
\begin{eqnarray}
&&\sigma_I^2[k]=\frac{1}{N}\sum_{l=0}^{L-1}\sigma_{\Delta h_l}^2 \left[ \sum_{n=0}^{\nu-1}|c[n]|^2\right. \nonumber\\
&+&\!\!\left. \sum_{q=1}^{\nu-1}2\cos\left(\frac{2\pi}{N}kq\right)\!\!\sum_{n=0}^{\nu-1-q}c_l[n]^{\ast}c_l^{\ }[n+q]\right]
\end{eqnarray}

\section*{Acknowledgment}
This work is carried out in the framework of the French research project ``Mobile TV World''.

\ifCLASSOPTIONcaptionsoff
  \newpage
\fi



%

\end{document}